\documentstyle[12pt]{article}
\begin{document}
\setlength{\textheight}{9in}
\setlength{\textwidth}{5in}
\baselineskip = 3mm

\def\theequation{\arabic{section}.\arabic{equation}}
\newcommand{\beq}{\begin{equation}}
\newcommand{\eeq}{\end{equation}}
\newcommand{\rbracket}{\right]}
\newcommand{\fff}{{\bar f}}
\newcommand{\lbracket}{\left[}
\newcommand{\DD}{{\cal D}}
\newcommand{\DDt}{{\nabla}_{t}}
\newcommand{\ddt}{\partial_t}
\newcommand{\ddx}{\partial_x}
\newcommand{\h}{\theta}
\newcommand{\pdv}{\partial_{\varphi}}
\newcommand{\r}{\cal G}
\newcommand{\Ag}{A^{g}}
\newcommand{\aaa}{{\alpha}}
\newcommand{\Q}{\bar{Q}}
\newcommand{\Pb}{\bar{P}}
\newcommand{\ppb}{\bar p_{i}}
\newcommand{\ps}{\bar{\psi}}
\newcommand{\LL}{{\cal L}}
\newcommand{\ggg}{g}
\newcommand{\RR}{{\cal R}_{\nu_{i}}}
\newcommand{\ep}{\epsilon_{i}}
\newcommand{\Vq}{\Delta (\Q)}
\newcommand{\Vp}{\Delta (\Pb)}
\newcommand{\rw}{\rightarrow}
\def\op{operator}
\def\tly{topologically}
\def\mfti{Moscow Institute of Physics and Technology}
\def\onlabs{on leave of absence from }
\def\itep{Institute of Theoretical and Experimental Physics}
\def\Clgr{Calogero}
\def\CG{Clebsh-Gordon}
\def\gl{global}
\def\an{anomaly}
\def\f{field}
\def\exst{existence}
\def\nl{normalizable}
\def\nty{normalizability}
\def\cn{condition}
\def\con{configuration}
\def\beq { \begin{equation}}
\def\eeq { \end{equation}}
\def\rw {\rightarrow}
\def\lw {\leftarrow}
\def\tt {\cal t}
\def\intt {\int_{S^{1}}}
\def\bbb {\box}
\def\lll {\lambda}
\def\ddd {\cal D}
\def\hh {\hat h}
\def\meas {{\ddd} A_{t} {\ddd} A_{x} {\ddd} \phi}
\setlength{\oddsidemargin}{-0.1in}
\title{ Relativistic Calogero-Moser Model as Gauged WZW Theory}
\author{\sf Alexander Gorsky
\thanks{e-mail: gorsky@vxitep.itep.msk.su}\\
\centerline{\em Institute of Theoretical and Experimental Physics }\\
\centerline{\em 117259, Bol. Cheremushkinskaya, 25, Moscow, Russia}\\
\centerline{\sf and}\\
 \sf   Nikita Nekrasov \thanks{permanent address:ITEP,
117259,B.Cheremushkinskaya, Moscow,
Russia }
\thanks{e-mail: nikita@rhea.teorfys.uu.se}\\
 \em Institute of Theoretical Physics at Uppsala University \\
 \em  Box 803  S-751 08 Uppsala Sweden}

\date{
\setlength{\unitlength}{\baselineskip}
\begin{picture}(0,0)(0,0)
\put(12,15){\makebox(0,0){UUITP-31/93}}
\put(12,14.4){\makebox(0,0){December 1993}}
\put(12,13.7){\makebox(0,0){hepth/9401017}}
\put(-10,-16){\makebox(0,0){ITEP-NG2/93}}
\end{picture}}

\maketitle
\abstract{We study  quantum integrable systems
of interacting particles from the point of view,
proposed in \cite{gor}. We obtain
Calogero-Moser
and Sutherland systems
as well as their
relativistic-like generalization due to Ruijsenaars
by  a hamiltonian
reduction of integrable systems on the cotangent bundles over semi-simple
Lie algebras, their affine algebras and central extensions of loop groups
respectively.  The corresponding 2d field theories form a tower of
deformations.  The top of this tower is the gauged G/G WZW model on a
cylinder with inserted Wilson line in appropriate representation.  Its
degeneration yields 2d Yang-Mills theory, whose small radius limit is
Calogero model itself.  We make some comments about the spectra and
eigenstates of the models which one can get from their
equivalence with the field theories. }
\newpage
\setcounter{equation}{0}

$\;\;\;\;$This paper should be considered as a natural continuation of
\cite{gor}. The question of determination of spectra and eigenstates in some
inegrable many-body systems will be addressed here. The main tool in our
approach will be a hamiltonian reduction approach to the systems with
symmetries. We shall consider two dimensional gauge theories, which can be
treated as a topological theories and due to their topological nature only
finite number of degrees of freedom form a "physical" Hilbert space of the
theories. This space can be interpreted as a state space in some quantum
mechanical problem. On the other hand, on can easily
calculate path integrals in the topological field theories under
consideration. It provides us with an information about the spectra and wave
functions in the quantum mechanical problems.

 We consider first the Sutherland model \cite{suth} at some distinguished
values of coupling constants and show that it can be obtained even at
quantum level as a result of hamiltonian reduction from the cotangent bundle
to the affine Lie algebra $\hat g$. Then we compare this reduction with the
reduction which is present in the two dimensional Yang-Mills theory and
realize
that these two in fact coincide. This permits us to get the
Harish-Chandra-like
formulas for the wavefunctions and for the kernel in the Sutherland model
(actually, we use here results of \cite{2dqcd}). This is
in the good relation with the idea of I.Cherednik to consider affine algebras
to get exact
answers for many-body integrable systems \cite{cherednik}. This result
generalizes results of \cite{2qcdpart}, while some methods of \cite{2qcdpart}
could be applied in our situation too to give a satisfying treatment of the
large $N$ behaviour of our systems.

Recently relativistic-like  Calogero-Moser-Sutherland models were found
\cite{ruu}. We show that these models (actually, their trigonometric version)
appear in the Hamiltonian reduction procedure applied to the $T^{*}{\hat G}$
(the continuation of approaches  of \cite{kag},\cite{olsh-per-cl}),
and that this reduction is equivalent to that in the   gauged   $G/G$
Wess-Zumino-Witten theory (and also in the Chern-Simons theory on a threefold
which is
a suspension of the two dimensional surface by a  circle). Thus, we are able
to tell something about the spectra and their degeneracy in these case.
To this end we need an expression for the path integral in the gauged $G/G$
WZW model on a disc. This expression can, in principle, be extracted from
\cite{bt + ger}.

\section{Affine Lie algebra approach to Sutherland model}
\setcounter{equation}{0}
\subsection{Hamiltonian reduction of $T^{*}{\hat g}$}
   Let us consider some semi-simple real Lie algebra $g$ with
the Killing form $<,>$ denoted also as a $\sf tr$ and let us fix an
isomorphism between $g$ and $g^{*}$ induced
from the $\sf tr$. Affine Lie algebra $\hat g$ is defined to be the central
extension of the loop algebra ${\cal L}g$ that is it is a space of the
pairs $({\phi}, c)$ where $\phi$ is a map from the circle $S^{1}$ to
the $g$ and
$c$ is a real number. The Lie algebra structure on $\hat g$ is introduced as
follows:

$$
[({\phi}_{1}({\varphi}),c_{1}), ({\phi}_{2}({\varphi}),c_{2})] =
([{\phi}_{1}({\varphi}),{\phi}_{2}({\varphi})],
\int_{S^{1}} <{\phi}_{1}, {\partial}_{\varphi} {\phi}_{2}>),
$$

Now let us turn to the dual space to the $\hat g$. This space ${\hat g}^{*}$
consists
of the pairs $(A, {\kappa})$
where $A$ is $g$-valued one-form (actually, a distribution) on
the circle and
$\kappa$
is just the real number. The pairing between the $\hat g$ and
${\hat g}^{*}$ is

$$
<(A, {\kappa}) ; (\phi , c)> = \intt <\phi, A> + c {\kappa}.
$$

The direct sum ${\hat g} \oplus {\hat g}^{*}$ we will denote as
$T^{*}{\hat g}$.
On this cotangent bundle the natural symplectic structure is defined:

\beq
\Omega = \intt {\sf tr} ( \delta \phi \wedge \delta A) + \delta c
\wedge \delta {\kappa}
\label{sympl}
\eeq

We can define adjoint and coadjoint actions of
the loop group ${\cal L}G$
on the $\hat g$ and ${\hat g}^{*}$. The former is defined from the commutation
relations and the latter from the pairing, that is the element
$g(\varphi)$
acts as follows:

\beq
(\phi(\varphi) , c) \rw (g(\varphi) \phi(\varphi) g(\varphi)^{-1},
\intt {\sf tr} (- \phi g^{-1}{\partial_{\varphi}}g) + c )
\label{adj}
\eeq

\beq
(A, {\kappa}) \rw (g A g^{-1} + {\kappa} g {\partial_{\varphi}} g^{-1},
{\kappa})
\label{coadj}
\eeq

This action clearly preserves the symplectic structure (\ref{sympl}) and thus
defines a moment map

$$
\mu : T^{*}{\hat g} \rw {\hat g}^{*}
$$

which sends a 4-tuple $(\phi, c; A, {\kappa})$ to the pair
$( {\kappa} d \phi + [A, \phi] , 0)$.

 Now let us choose an appropriate level of moment map and make a hamiltonian
reduction under this level. To guess, what element of ${\hat g}^{*}$ we should
choose it is convinient to return back to the finite-dimensional example
of this
procedure. Namely, if one considers a reduction of the $T^{*}g$
with respect to
the action of $G$ by conjugation, then, to get a desired Calogero system,
one takes
an element $J$ of $g^{*}$, which has maximal stabilizer, different from
the whole $G$.
It is easy to show, that the representative of the coadjoint orbit of this
element has the following form:

\beq
J_{\nu} = {\nu} \sum_{\alpha \in \Delta_{+}}( e_{\alpha} + e_{-\alpha}),
\label{mom}
\eeq

where $\nu$ is some real number,  $e_{\pm \alpha}$ are the elements of
nilpotent subalgebras $n_{\pm} \subset g$, which correspond to the roots
 $\alpha$, and ${\Delta}_{+}$ is the set of positive roots. In fact, if the
corresponding Coxeter group has more then one orbit in the set of roots (i.e.
two)
then there are more coupling constants \cite{dunkl}.
Let us denote the coadjoint orbit of $J_{\nu}$ by ${\cal O}_{\nu}$
and by the $R_{\nu}$
the representation of $G$, arising upon the quantization of ${\cal O}_{\nu}$.
For generic $J \in g^{*}$ let
$G_{J}$ will denote the stabilizer of $J$ in the coadjoint orbit
${\cal O}_{J}$
of $J$, i.e. ${\cal O}_{J} = G / G_{J}$.

 Now let us return back to the case of affine Lie algebra. We could just take
the formula (\ref {mom})
and replace everywhere usual roots by affine ones, but we prefer to have more
geometrical reasoning.
Because of the vanishing level ${\kappa}_{\mu}$ of the moment map,
the coadjoint orbit of generic value of
it is rather huge, while the stabilizer of the element
$(J({\varphi});0) \in {\hat g}^{*}$
is a "continious product" $\prod_{{\varphi} \in S^{1}} G_{J({\varphi})}$
which is very small
in comparison with the whole loop group ${\cal L} G$ unless for almost all
${\varphi} \in S^{1}$ the group $G_{J({\varphi})}$ coincides with $G$.

 All this argumentation shows that an appropriate value of
 the moment map we are
looking for is the following:

\beq
\mu = ({\cal J}[{\mu}],0) : {\cal J}[{\mu}]({\varphi}) = \delta({\varphi})
J_{\nu}
\label{affmom}
\eeq

 Here by $\delta ({\varphi})$ we mean a distribution ( one-form)
supported at some marked point $0 \in S^{1}$ and normalized in such a way,
that
 $\int_{S^{1}} \delta ({\varphi}) = 1$.
The coadjoint orbit of $\mu$ is nothing, but the finite-dimensional orbit
${\cal O}_{\nu}$.

 Now let us complete the reduction. To this end we should resolve the equation

\beq
{\kappa} \partial_{\varphi} \phi + [A, \phi] =  J_{\mu}({\varphi}) =
\delta({\varphi}) J_{\nu}
\label{redeq}
\eeq

modulo the action of stabilizer of $\mu$, that is modulo subgroup of
${\cal L}G$,
consisting of $g({\varphi}) \in G$, such that $g(0) \in G_{\nu}$. We can do
it as follows.
First we use generic gauge transformation $\tilde g({\varphi})$ to make $A$
to be the Cartan subalgebra $t \subset g$
-valued constant one-form $D$ (it is always possible for non-vanishing
$\kappa$). We are left with the
freedom to use the constant gauge transformations with values in
the Cartan subgroup
${\bf T} \subset G$, generated by $D$ - these do not touch $D$.
Actually, the choice of $D$ is not unique. The only
invariant of ${\cal L}G$ action is  the conjugacy class of monodromy
$\exp ( \frac{2\pi}{\kappa} D) \in {\bf T} \subset G$. Let us fix some of
these choices. Let
${\sf i} x_{i}$ will denote the entries of $D = iX$. Let us decompose
the $g$-valued function
$\phi$ on the $S^{1}$ on its Cartan-valued part $P(\varphi) \in t$
and let ${\phi}_{\pm}({\varphi}) \in n_{\pm}$ be its
nilpotent-valued parts. Let ${\phi}_{\alpha} = <{\phi}, e_{\alpha}>$.
 Then the equation (\ref{redeq}) will take the form:

\beq
{\kappa}\partial_{\varphi} P = \delta({\varphi}) [ J_{\nu}^{g} ]_{\gamma}
\label{careq}
\eeq

\beq
{\kappa}\partial_{\varphi}{\phi}_{\alpha} + <D, {\phi}_{\alpha}> =
\delta({\varphi}) {[ J_{\nu}^{g} ]}_{\alpha}\\
\label{rooteq}
\eeq

where $J_{\nu}^{\tilde g}$ is simply $Ad_{{\tilde g}(0)}^{*}(J_{\nu})$,
$[J]_{\gamma}$ denotes the Cartan's part of $J$ and

$[J]_{\alpha} = <J,e_{\alpha}>$.

 From the equation (\ref{careq}) we deduce that $D = constant$ and
 $[J_{\nu}^{g}]_{\gamma} = 0$.
This implies, that be Cartan-valued constant conjugation we can twist
$J^{g}_{\nu}$ to the $J_{\nu}$ itself.
 Then, (\ref{rooteq}) implies that locally (at $\varphi \neq 0$ )
we can represent ${\phi}_{\alpha}({\varphi})$ as follows:

\beq
{\phi}_{\alpha}({\varphi}) =
\exp ( - \frac{\varphi}{\kappa} <D, {\alpha}> ) \times M_{\alpha}
\label{rootres}
\eeq

 where $M_{\alpha}$ is locally constant vector in $g$. Look at the right hand
side of (\ref{rooteq}) leads us to the conclusion, that $M_{\alpha}$ jumps
when ${\varphi}$ goes through $0$. This jump is equal to

\beq
[\exp ( - \frac{2\pi}{\kappa} <D, {\alpha}> ) - 1] \times M_{\alpha}
= [J^{g}_{\nu}]_{\alpha}
\label{jump}
\eeq

The final answer for the reduction is the following:
the physical degrees of freedom are in the
$\exp(-\frac{2\pi i}{\kappa}X)$
and $P$ (note that $P$'s entries are purely imaginary) with the reduced
symplectic structure:

\beq
\omega = \frac{1}{2 \pi i} {\sf tr} ( \delta P \wedge \delta X )
\label{redsym}
\eeq

and we have

$$
{\phi}_{\alpha}({\varphi}) = {\nu}
\frac{\exp ( - \frac{i\varphi}{\kappa} <X, {\alpha}> )}
{\exp ( - \frac{2\pi i}{\kappa} <X, {\alpha}> ) - 1}
$$

Now if we would take some simple hamiltonian system on the $T^{*}{\hat g}$
whose
Hamiltonian is invariant under (\ref{adj}),(\ref{coadj}) actions, then we will
get somewhat complicated system on the reduced symplectic manifold, i.e. on
$T^{*}{\bf T}$.
 For example, let us take any invariant polynomial $Q(\phi)$ on
 the Lie algebra $g$,
that is $Q \in S^{\cdot}(g^{*})^{G} = S^{\cdot}(t^{*})^{W}$, where $W$ is
the Weyl group of
$G$. Then we can construct a Hamiltonian on $T^{*}({\hat g})$ by the formula

$$
{\cal H}_{Q} = \frac{1}{2\pi} \int_{S^{1}} d{\varphi} Q(\phi)
$$

with somehow fixed one-form $ d{\varphi}$ (for example,
it will be convinient for us to fix
a normalization $\frac{1}{2\pi} \int_{S^{1}} d{\varphi} = 1$). For quadratic
casimir we get

\beq
{\cal H}_{2} = \frac{1}{4\pi} \int_{S^{1}} d{\varphi} <{\phi}, {\phi}>
\label{2cas}
\eeq

On the reduced manifold we are left with the Hamiltonian

\beq
H_{2} = -\frac{1}{2} {\sf tr} P^{2} +
\sum_{\alpha \in \Delta_{+}} \frac{{\nu}^{2}}
{sin^{2}<X,{\alpha}>}
\label{redham}
\eeq

which coincides with the Hamiltonian of the Sutherland model.
Here $\Delta_{+}$ denotes the set of positive roots
of $g$. For example, for $G = SU(N)$ we have the Hamiltonian
for pair-wise interaction between
the particles with the potential:
$$
V_{ij}^{A} = \frac{g^{2}}{sin^{2}(x_{i}-x_{j})}
$$

for $G = SO(2N)$ we have:
$$
V_{ij}^{D} = g_{2}^{2} [ \frac{1}{sin^{2}(x_{i}-x_{j})} +
\frac{1}{sin^{2}(x_{i}+x_{j})}] +
$$
$$
+ g_{1}^{2} [\frac{1}{sin^{2}(x_{i})} + \frac{1}{sin^{2}(x_{j})}]
$$

where $g_{1}, g_{2}$ are coupling constants. In general
case one has as many different coupling constants as many
orbits in
the root system has a Weyl group, i.e. two or one \cite{olsh-per-cl}.

\subsection{Two-dimensional Yang-Mills theory}

Let us consider some
two dimensional surface $\Sigma$ with fixed non-degenerate
two-form $\omega\in\Omega^{2}(\Sigma)$ but
without fixed metrics. Then one can consider some principal
$\sf G$-bundle $\cal E$ over $\Sigma$ and the space $\cal A$ of connections
on $\cal E$.
This space is a symplectic one
and group $\cal G$ of gauge transformations acts on it preserving that
symplectic structure.
The moment map for this action in fact coincides with the curvature
$F = dA + A \wedge A$
of the connection $A$ unless our surface $\Sigma$ has holes.
In that case one has to specify what
are boundary conditions for the gauge transformations.

Let us take a surface $\Sigma$ with some fixed contours $C_{\beta}$
(in the case of
holes one can consider non-closed contours with the ends on this holes).
Then for each loop $C$
a natural gauge invariant observable is defined :
$W_{C} = \chi_{\alpha}(P\exp \oint_{C}A ) $ where $\alpha$ is some
irreducible representation of $G$. For non-closed contours
one should attach to each end of it a representation of $G$.
Contours should be oriented
and representations ${\alpha}_{\Gamma},{\beta}_{\Gamma}$, belonging
to the ends of the same contour $\Gamma$ are to be dual. Then,
the observable we wish to construct takes its values in the tensor product
${\alpha}_{\Gamma}
\otimes {\alpha}_{\Gamma}^{*}$ (where ${\alpha}_{\Gamma}$ belongs to that
end from this this contour departs due to orientation) - it is the group
element $P \exp \int_{\Gamma} A \in G$
in the representation ${\alpha}_{\Gamma}$, i.e.
$W_{\Gamma} = T_{{\alpha}_{\Gamma}}(P \exp \int_{\Gamma} A)$.
This is an example
of the coloring of the (fat) graph in terminology of \cite{fro}.

Now let us write down the action of Yang-Mills-like gauge theory on $\Sigma$.
 To this end we need an
extra field
$\phi \in \mbox{Maps} ({\Sigma}, {\sf g})$.

Let us take any invariant polynomial
function
on $g$, i.e.
$Q \in S^{\cdot}(g^{*})^{G}$. Then the action is:

$$
S_{\Sigma, Q} =
$$
\beq
= \int_{\Sigma}  < \phi, F> + \omega Q(\phi)
\label{2daction}
\eeq

Let us locally decompose the tangent bundle to $\Sigma$ to the time-like and
space-like directions.
This is a
choice of real polarization - it has some difficulties when the surface has
pants-like vertices.
In the vicinity of generic point $P \in \Sigma$ we can always find
such a coordinates $t,x$ that 2-form $\omega$ will take its canonical form;
$$
\omega = dt \wedge dx
$$

We suppose
the orbits of the vector field $\frac{\partial}{\partial x}$ to be periodic
almost everywhere.
The symplectic structure on $\cal A$ (more precisely, that part of it,
which corresponds to the vicinity of $P$) is

$$
\Omega = \int_{\Sigma} <\delta A_{t}, \delta A_{x} > dt \wedge dx
$$

(This symplectic structure, certainly, doesn't depend on the $\omega$ or
whatever,
it is simply $\Omega = \int_{\Sigma} {\sf tr} (\delta A \wedge \delta A)$.)

Now let us suppose, that some contour $\Gamma$ goes straight along
the time-like direction
$\frac{\partial}{\partial t}$. Then we get the following path integral
(we disregard
now any kinds of $\Sigma$ except spheres with two holes
$\gamma_{1}, \gamma_{2}$,
and on each hole points $P_{1}$ and $P_{2}$ respectively are marked.
Contour $\Gamma$ goes
from $P_{1}$ to $P_{2}$):

$$
\int \meas \exp [ - S_{S^{1} \times I, Q}]
<v_{1}| P \exp \int_{\Gamma} A |v_{2}>
$$

here $<v_{1}|, |v_{2}>$ are the vectors in the representations
${\alpha}_{\Gamma}, {\alpha}_{\Gamma}^{*}$ respectively.
We wish to take
${\alpha}_{\Gamma}$ to be equal $R_{\nu}$ from the previous section.

Let us remind some facts about $<v_{1}| P \exp \int_{\Gamma} A |v_{2}>$.
It follows ideologically the Kirillov constructions of the representations
(and it was checked in \cite{samson}),
that this
matrix element can be represented as a path integral with boundary conditions.
We have to take a coadjoint orbit ${\cal O}_{\nu}$ which corresponds to the
$R_{\nu}$ choose one-form $\theta = d^{-1} \omega_{\cal O}$, where
$\omega_{\cal O}$ is
a natural Kirillov's form on ${\cal O}_{\nu}$ and then we can write
a desired path integral:

$$
<v_{1}| P \exp \int_{\Gamma} A |v_{2}> \sim \int {\ddd}{\chi} \exp [-\int
(\theta - <A_{t} , {\mu}_{\cal O}>)]
$$

where integral in
the r.h.s. needs some boundary conditions, which can be specified after
choosing a
polarization on the orbit
${\cal O}_{\nu}$, ${\mu}_{\cal O}: {\cal O}_{\nu} \rw g^{*}$
is a moment map corresponding to the natural $G$ action on the orbit.
 Finally, ${\ddd}{\chi}$
 is a path integral measure, constructed from the $\omega$.
In the case
we are interested in,
(and for $G =SU(N)$) ${\cal O}_{\nu} = {\bf CP}^{N-1}$ with the
symplectic form
$\omega_{\cal O} = N{\nu} \times$ Fubini-Shtudi form.
The corresponding moment map
${\mu}_{\cal O}$ sends
the point $(z_{0}:z_{1}: \dots :z_{N-1}) \in {\bf CP}^{N-1}$
to the matrix  $J_{ij} = \nu
(\delta_{ij} - z_{i}{\overline z_{j}}) \in su(N)^{*}$. So, we can rewrite
the matrtix element of the Wilson line as the following path integral:

\beq
\int {\ddd}z_{i}{\ddd}{\overline z_{i}}
\exp [ -\nu \int \sum_{i} {\overline z_{i}}\partial_{t} z_{i}
 - \sum_{ij} A_{ij}(t) z_{i}{\overline z_{j}}]
\label{wils}
\eeq

Then,  we put this expression into the Yang-Mills path integral and
see, that the field $A_{t}$  plays the role of Lagrangian multiplier,
giving rise to the constraint (\ref{redeq}).
Resolving it (modulo gauge transformations, preserving
the point on the coadjoint orbit) we  get the same finite-dimensional
Hamiltonian for the eigenvalues of the monodromy around the space-like slice,
as we had in the previous section. This proves the equivalence between these
two theories.

We see, that the whole construction to work, we need the orbit
${\cal O}_{\nu}$ to be quantizable. This implies (in the case of $G = SU(N)$)
that $N \nu$ should be integer.
Further we will get more restrictive condition.

\subsection{Wavefunctions and amplitudes for the Sutherland model}
What can we get from this equivalence? We can solve the Sutherland model
due to
possibility of simple solving of the 2d Yang-Mills theory. Namely, in 2d YM
theory,
we can calculate exactly the path integral on the discomputed with the help of
the form $\omega$ and
$a_{\ddd}$ is the area of ${\ddd}$,i.e. $\int_{\ddd} \omega$.

In our situation we have to attach a representation $R_{\nu}$ to the
outgoing end of the Wilson line and the dual to the incoming one.
So, the boundary conditions we
should fix are: vectors $<v_{1}| \in R_{\nu}^{*}, |v_{2}> \in R_{\nu}$ and the
monodromies
$ g_{2} = P\exp \int_{0_{2}} A$, $ g_{1} = P\exp \int_{0_{1}} A $ around
the initial and final holes respectively.

Gauge transformations will rotate
monodromies and vectors as follows:
$g_{i} \rw h g_{i} h^{-1}, |v_{i}> \rw T_{R_{\nu}}(h) |v_{i}> \;
i = 1,2$ and the answer will be invariant with
respect to these transformations.
The measure on $g_{i}$ is the Haar measure. Using this
gauge freedom, we can make $g$ to be in the Cartan subgroup ${\bf T} \in G$.
The measure
on ${\bf T}$ induced from the modding out angle variables is
the product of the
Haar measure on the ${\bf T}$ - $\prod_{i} d\theta_{i}$ and corresponding
group-like VanderMonde determinant.
To get the proper answer for the transition amplitude we should
take a square root of this measure (to get a half-form).
Finaly, note that in the Sutherland model itself we have a quadratic
$Q(\phi) = \frac{1}{2} {\sf tr} ({\phi}^{2})$.

 Now let us turn to the calculation. Let us cut our cylinder along
 the contour $\Gamma$:
we will get the disk, and on the edges, corresponding to $\Gamma$ we have
some group element $h \in G$, which should be integrated out with the weight
$<v_{1}|T_{R_{\nu}}(h)|v_{2}>$.
 The whole monodromy around the disk is given by $g_{1}hg_{2}^{-1}h^{-1}$,
so the answer for the transition amplitude (or evolution kernel) is
the following:
\beq
<g_{2};v_{2}|\exp[-tH]|g_{1};v_{1}> =
\label{kernel}
\eeq
$$
= \sum_{\hh} d_{\hh} e^{-tQ_{2}(\hh + \rho)} \int dh
{\chi}_{\alpha_{\hh}}(g_{1}hg_{2}^{-1}h^{-1}) <v_{1}|T_{R_{\nu}}(h)|v_{2}>
$$

{}From this expression we can extract the spectrum of the model as well as the
structure of the Hilbert space (this can be achieved also by looking at the
$T^{*}G$ \cite{poly}):

$$
{\cal H} = \bigoplus_{\hh} Inv (\alpha_{\hh} \otimes {\alpha_{\hh}}^{*}
\otimes R_{\nu}) =
$$
$$
= \bigoplus_{\alpha} \sum_{\Phi_{\hh}} {\bf C} \cdot \Phi_{\hh}
$$
where $\Phi_{\hh}: \alpha_{\hh} \to \alpha_{\hh} \otimes R_{\nu}$
is an intertwiner.

$$
E_{\alpha_{\hh}} = \frac{1}{2}<\hh + \rho, \hh + \rho>
$$

{}From the answer for the kernel we can realize that more restrictive
conditions on coupling $\nu$ should
be imposed to be possible to get the Sutherland system
as quantum reduction:
if we take $g_{1},g_{2}$ to be ${\bf T}$ - valued, then the integral
over $h$ will invariant
under left and right independent multiplication of $h$ by
the elements of ${\bf T}$ -
this lead to the condition $T_{R_{\nu}}({\bf T})|v_{i}> = |v_{i}>$. It is the
integral $\nu$, for which $R_{\nu}$ contains such a vector (vacuum vector
of $R_{\nu}$).
Moreover, there exist only one such a vector. Let us denote it as $|0>$.

To get the expression for the wavefunction, let us look once again on the
integral over $h$.
It can be taken by representing character in the orthonormal base of
representation $\alpha_{\hh}$,
so we will arrive to the expression:

\beq
\Psi_{\alpha_{\hh}}( \{ \theta_{i} \} ) = {\cal N} (\{ \theta_{i} \})
\sum_{mn} C_{mn} <m|T_{\alpha_{\hh}}(diag[e^{{\sf i}\theta_{i}}])|n>
\label{wf}
\eeq

here ${\cal N} (\{ \theta_{i} \})$ denotes the normalization factor,
which comes
from
the taking into account the group-like Vandermonde determinant,$m,n$ run over
the orthogonal base
of the representation $\alpha$  and coefficients $C_{mn}$ are defined via:

$$
\int dh T_{\alpha}(h)_{mk} T_{\alpha}(h)_{nl}^{*} T_{R_{\nu}}(h)_{00} =
C_{mn} C_{kl}^{*}
$$

so they are just Littlewood-Richardson (or Clebsh-Gordan) coefficients.

It is clear, that this consideration generalizes that of \cite{2qcdpart}.


\section{Deformation of Calogero-Moser models - trigonometric Ruijsenaars's
system.}
\subsection{Hamiltonian reduction of the cotangent bundle of a central
extension of loop group}
\setcounter{equation}{0}

In this section we will consider the deformation of the previous
constructions.
Namely, let us consider the Hamiltonian reduction of the cotangent bundle to
the central extended loop group $\hat G$. This symplectic manifold is a
space of 4-tuples

$(g:S^{1} \rw G, c \in U(1) ; A \in \Omega^{1}(S^{1}) \otimes g^{*}, {\kappa}
\in {\bf R})$.

The action of an element $h \in {\cal L}G$ is the following:

$$
g \rw h g h^{-1}, \; A \rw h A h^{-1} + {\kappa} h {\partial} h^{-1}
$$

$$
{\kappa} \rw {\kappa}, \;
c \rw c \times {\cal S}( g, h)
$$

where $\cal S$ is constructed from the $U(1)$-valued two-cocycle on the group
${\cal L}G$,
$\Gamma (g, h)$
(it provides ${\cal L}G$ with a central extension)
(the cocycle is supposed to be normalized : $\Gamma (g, g^{-1}) = 1$):

$$
{\cal S} (g, h) = \Gamma (h, g) \Gamma(hg, h^{-1})
$$

As always, on this cotangent bundle exists a natural symplectic structure and
the
loop group action preserves it. Explicitely
this form $\Omega$ can be written in
some trivialization of the cotangent bundle $T^{*}{\hat G}$. Namely, we
identify it with the direct product ${\hat {\sf g}} \times {\hat G}$
with the help of left-invariant one-forms ${\kappa}{\partial_{\varphi}} +
A_{\varphi} \in \Omega^{1}({\hat G})$. The form $\Omega$ has the
following structure:
\begin{eqnarray}
 && \Omega = \int_{S^{1}} {\sf tr}
[
A (g^{-1} \delta g)^{2}
+ \delta A \wedge
g^{-1} \delta g ] +
\nonumber\\
&& \int_{S^{1}} {\sf tr} [ {\kappa}
{\pdv}g \cdot g^{-1} (\delta g \cdot g^{-1})^{2} -
\kappa \delta ({\pdv}g) \cdot g^{-1} \delta g \cdot g^{-1}] +
\nonumber\\
&& + c^{-1} \delta c \wedge \delta \kappa
\label{loopsympl}
\end{eqnarray}

The moment map of the action of loop group
has the form:

\beq
\mu (g,c; A,\kappa) = (g A g^{-1} + {\kappa} g {\partial} g^{-1} - A, 0)
\label{momgru}
\eeq

It is tempting to make the Hamiltonian reduction under some appropriate
level of
the moment map. It is clear that in this way we deform Sutherland system,
presumably preserving its integrability. For simplicity we consider only
$SU(N)$ case.
The reasoning, similar to that in the YM case, suggests
the level of moment map to be

\beq
\mu (g,c; A,{\kappa}) = {\sf i} \nu (\frac{1}{N}
Id - e \otimes e^{+}) \delta (\varphi)
\label{momlev}
\eeq

As we did it in the previous sections, by general gauge transformation
$H(\varphi)$
we can make $A$ to be constant diagonal matrix $D$, defined modulo affine
Weyl group action, i.e. modulo permutations and addings of the integral
diagonal matrices. Generally, such a transformation
doesn't respect the value of the moment map. Let $H$ denotes the value of
$H(0)$
at the point $0$. The matrix $H$ defines a point in the coadjoint orbit of
$J$,
i.e. ${\cal O}_{J} ={\bf CP}^{N-1}$. The semi-direct product
of Cartan torus $U(1)^{N-1}$ and Weyl group ${\cal S}_{N}$ acts on
${\cal O}_{J}$ by permutations of homogeneous coordinates and
their multiplications by phases. The quotient ${\cal O}_{J} / U(1)^{N-1}$
coincides with $(N-1)$- simplex.

The equation \ref{momgru} takes the form:

\beq
g D g^{-1} + {\kappa} g d g^{-1} - D =
{\sf i} \nu (\frac{1}{N} Id - f \otimes f^{+})\delta(\varphi)
\label{mastereq1}
\eeq

Here $f = H e$ is some vector in ${\bf C}^{N}$ with unit norm $<f,f> = 1$.
We can assume that $f \in {\bf R}^{N}$, due to the remaining possibility of
the left multiplication $H \rw Y H$ with $Y$ being diagonal unitary matrix
(it comes from the gauge freedom which survives after diagonalization of
$A$). Equation (\ref{mastereq1}) yields immediately:

$$
g = \exp ( \frac{\varphi}{\kappa} D) G(\varphi)
\exp( -\frac{\varphi}{\kappa} D) ;
\partial_{\varphi} G = - \frac{J}{\kappa} G \delta(\varphi) $$

where $J = {\sf i} \nu ( \frac{1}{N} Id - f \otimes f^{+} )$.
Let us also introduce a notation for the monodromy
of connection $D$: $Z = \exp ( - \frac{2\pi}{\kappa} D) = diag(z_{1}, \dots,
z_{N})$,  $\prod_{i} z_{i} = 1$,
$z_{i} = \exp (\frac{2{\pi}i q_{i}}{\kappa})$.
We have a boundary condition:

\beq
 {\tilde G}^{-1} Z {\tilde G} = \exp(\frac{2\pi J}{\kappa}) Z
\label{commutant}
\eeq

where $\lll = e^{\frac{2 \pi i\nu}{N \kappa}}$,$ {\tilde G} = G(+0)$. Though
the resolution of this equation is rather interesting issue, we postpone it
till the Appendix A. It turns out that solution has the following form: let
$P(z)$ will denote a characteristic polynomial of $Z$, i.e.  $$ P(z) =
\prod_{i} (z - z_{i}) $$

Let
$$
Q^{\pm}(z) =
\frac{P({\lll}^{\pm 1}z) - P(z) }{({\lll}^{\pm N}-1)z P^{\prime}(z)}
$$
- the ratio of the finite difference and derivative of $P$.
When $\lll \to 1$,
the rational functions $Q^{\pm}(z)$ tend to $\frac{1}{N}$.

In these notations the matrix $\tilde G$ can be written as follows:

\beq
{\tilde G}_{ij} = - \lll^{-\frac{N-1}{2}}
\frac{{\lll}^{-N}-1}{{\lll}^{-1}z_{i}-z_{j}}
e^{i\theta_{i}}(Q^{+}(z_{i})Q^{-}(z_{j}))^{1/2}
\label{Lax}
\eeq

$$
=e^{i\h_{i}-\frac{\pi i}{\kappa} (q_{i}+q_{j}) }
\frac{sin(\frac{\pi\nu}{\kappa})}{sin(\frac{\pi(q_{ij}-
\frac{\nu}{N}}{\kappa})}
\prod_{k \neq i, l \neq j}
\frac{sin(\frac{\pi{q_{ik} +
\frac{\nu}{N}}}{\kappa})}{sin(\frac{{\pi}q_{ik}}{\kappa})}
\frac{sin(\frac{{\pi}{q_{il} -\frac{\nu}{N}}}{\kappa})}{sin(
\frac{{\pi}q_{il}}{\kappa})}
$$

where
$\h_{i}$ are some
undetermined phases,
which are going to be the momenta,
corresponding to the coordinates  $q_{i}$.
It means, that on the reduced symplectic manifold the
symplectic structure is $\sim \sum_{i} d\h_{i} \wedge dq_{i}$.

The next step is to consider some simple Hamiltonian system on
$T^{*}{\hat G}$, which is invariant with respect to the loop group adjoint
action and make a reduction. It is obvious that any $Ad$-invariant function
$\chi : G \rw {\bf R}$ defines a Hamiltonian

$$
H_{\chi} = \intt d{\varphi} {\chi}(g)
$$

For example, $\chi_{\pm} (g) = {\sf Tr} (g {\pm} g^{-1})$ (here $\sf Tr$
is a
trace
in the $N$-dimensional
fundamental
representation of $SU(N)$ )
give the Hamiltonians (up to a constant factor, independent of $q_{i}$) :

\beq
H_{\pm} = \sum_{i} (e^{i \h_{i}} {\pm} e^{- i\h_{i}})
\prod_{j \neq i} f(q_{ij})
\label{ruuham}
\eeq

where the function $f(q)$ is given by:
$$
f^{2}(q) = [ 1 -
\frac{sin^{2}
({\pi}{\nu}/{\kappa} N)}
{sin^{2}({\pi}q/{\kappa})}
].
$$

For detailed elaboration of different limits of the Hamiltonian
see, for example \cite{ruu}.  In particular, taking an appropriate scaling
limit $\kappa \to \infty$ will return us back to Sutherland model.
Note also, that in order to have a real-valued Hamiltonian we need
$\frac{\nu}{N} < q_{ij}$
for any $i{\neq}j$,
thus $\frac{\nu}{\kappa}$ should be
restricted from above.

\subsection{ On Gauged $G/G$ WZW Model}

Starting from the very
symplectic form (\ref{loopsympl}) and the moment map (\ref{momgru})
we can write a geometric bulk action for the field theory.
It will be a sum of the term like $\int pdq$ and a moment value fixing
term $\int A_{t} \mu dt$, where $A_{t}$ is going to be a time-like
component of a gauge field, serving as a Lagrange multiplier.
Thus, the action is (we omit the term $c^{-1}{\partial_{t}}c \kappa$,
while fix the non-vanishing level $\kappa$):

$$
S( A, g)
= \int d{\varphi}dt
{\;\sf tr} [
- A_{\varphi} g^{-1} {\partial_{t}} g
- \kappa {\partial_{t}}g \cdot g^{-1} \cdot {\pdv} g \cdot
g^{-1}
 +
$$

$$
+ \kappa d^{-1}
({\pdv} g \cdot g^{-1} (dg \cdot
g^{-1})^{2}) +
$$

$$
+ A_{t} ( \kappa g^{-1}
{\pdv} g + g^{-1} A_{\varphi} g -
A_{\varphi} )]
$$

This is the action of $G/G$ gauged WZW model \cite{gwzw}, a topological
gauge theory recently shown to give a Verlinde formula for the dimension of
the space of conformal blocks in WZW theory on the surface \cite{bt + ger}.

Our theory will contain also a Wilson line in the representation
$R_{\nu}$. To compute the path integral on a cylinder, as in the case
of Yang-Mills theory,one cuts a cylinder along the Wilson line and the
answer for the evolution cernel will have the form (\ref{kernel}).In fact
the representations will run only through integrable representations of
$SU(N)_{k}$. We conjecture that the answer can be obtained also from
$U_{q}(SL_{N})$ representation theory\cite{ek}, thus establishing the
relation between this quantum group and gauged WZW theory once again.

\subsection{Chern-Simons interpretation}

Now we present the interpretation of the previous section construction
in terms of
Chern-Simons field theory,
deformed by some Hamiltonian. Let us consider a
Chern-Simons theory with gauge group $SU(N)$ on the space-time manifold
 being a product
of the interval and a two-torus $X = I \times T^{2}$. Its action is given by:

\beq
S_{CS} = \frac{i {\kappa}}{ 4\pi}\int_{X} {\sf tr} (A \wedge dA +
\frac{2}{3} A \wedge A \wedge A)
\label{CS}
\eeq

We know that the phase space of Chern-Simons theory on a
${\Sigma} \times I$ with
$I$ being an interval is a moduli space of flat connections. In the case of
inserted into the path integral Wilson lines, let us say just one Wilson
line in the representation $R$,  when the integral in hands looks like:

$$
{\ddd} A <v_{1}| T_{R}(P \exp \int A ) |v_{2}> \exp( - S_{CS}(A) )
$$
the answer will be slightly changed, namely we will have to consider a
moduli space of connections on the punctured surface
with prescribed conjugacy class of monodromy of connection
around the puncture. This class $U$ is
 related to the highest weight $\hat h$ of the representation $R_{\nu}$ like
$ U = \exp ( \frac{2\pi {\sf i}}{{\kappa} + N} \mbox{diag} ( {\hat h}_{i}))$
If our surface is a two-torus $\Sigma = T^{2}$, then we can write
immediately the condition on the monodromies $g_{A}, g_{B}$ of connection
around $A$ and $B$ cycles and the monordomy $g_{C}$ around the puncture :
$$
g_{A} g_{B} g_{A}^{-1} g_{B}^{-1} = g_{C}
$$
Keeping in mind
the condition on the conjugacy class of $g_{C}$, taking into
account that for the representation $R_{\nu}$ we have a sighature like ${\nu}
\times \mbox{diag}
(N, 0, \dots, 0)$ we arrive to the equation
(\ref{commutant}).
The (non-rigorous) explanation of this relation between the
connections on
the two-torus and $T^{*}{\hat {\sf G}}$ is the following. One
can consider a hamiltonian
reduction of the symplectic space of connections on
the two-torus
by the subgroup of the group of gauge transformations, which are
equal to
identity at some (non-contractible) contour $C$ on the torus. If one
chooses a singular co-adjoint orbit of  the moment map $F = B \delta_{C}$,
where $B$ is a one-form on the $C$ and
$\delta_{C}$ is a delta-function in
the direction , normal to the contour $C$,
i.e. the connection
$A$ is allowed to have a jump at the $C$, then after a
reduction one is left with the limiting values of connection $A_{+}$ and
$A_{-}$ and with a ${\sf G}$-valued function on
$C$ - monodromy $g(x)$ ($x$ is
a coordinate on $C$) of connection around the cycle which begins and ends at
the point $x \in C$ and which is
transversal to $C$ in homology $H^{1} (T^{2},
{\bf Z})$.
The boundary values $A_{+}$ and $A_{-}$ are related via the gauge
transformation $g(x)$, i.e.
$A_{+}^{g} = A_{-}$\footnote{We are grateful to
A.Alexeiev for a discussion on this point}.

 More rigorous,
but less transparent
explanation involves a consideration of the space of
connections
on the annulus,
where the gauge group acts with a non-trivial cocycle in a Poisson
brackets of its Hamiltonians.
This cocycle  vanishes on a "diagonal"
subgroup, which is a group of gauge transformations on the annulus,
whose boundary values are related by some diffeomorphism $\varphi$ of
one component
of the
boundary onto another, i.e. $g(x) = g(\varphi(x))$, with $x$
being a point on the
one component and $\varphi(x)$ is its image in the other. Then
it is easy to see, that the hamiltonian reduction of the space of connections
by the action of this group can be done in two steps; first, we reduce
by gauge transformations, equal to identity at the boundary of the annulus,
then we take a symplectic quotient with respect to this "diagonal" subgroup,
isomorphic to the loop group ${\cal L}{\sf G}$\footnote{
We are grateful to V.Fock for valuable discussions on this point }.

Also we could point out the
equivalence of
the sector of the space of
observables in the
Chern-Simons theory on
${\Sigma} \times S^{1}$ ,
namely, observables,
which depend on
$A_{t}$ with ${\partial}_{t}$
being a direction, tangent to $S^{1}$ and the gauged $G/G$ WZW theory.
This can be proved either considering a gauge fixing
$A_{t} = \mbox{const}_{t}$ and diagonal,
or by poining out that
the wavefunctions in
the Chern-Simons theory are
those in the theory with an
action
$\int dt \int_{\Sigma} {\sf tr} (A {\partial_{t}} {\bar A})$
projected onto the space of functionals,
invariant under the gauge group action.
This involves taking into
account the non-invariance of
the polarization
$\frac{\delta}{\delta{\bar A}}$
and this is why the
WZW part of
the action appears (see \cite{bt + ger} for detailed discussion).

\subsection{The spectrum and wavefunctions of the model}

{}From this equivalence we can derive a spectrum of the
Ruijsenaars model.
Namely we will show that it is contained in the spectrum of
the free relativistic fermionic system on the circle.
By free relativistic
model of fermions on
the circle we
mean the system of
particles on
the circle with coordinates
$0 < q_{i} \leq 1$ and conjugate momenta
(rapidities) ${\h}_{i}$ living also on the circle, with the Hamiltonian
$$
H_{+} = \sum_{i} cos (\h_{i})
$$
and symplectic structure
$$
\omega =\frac{1}{2\pi} ({\kappa} + N) \sum \delta \h_{i} \wedge \delta q_{i}
$$
(in fact, the shift ${\kappa} \to {\kappa} + N$
is a quantum effect
and can be deduced from
\cite{bt + ger}, so we don't consider its derivation here).
The
fermionic nature of these particles
amounts to the phase space
$({\bf T}^{N-1} \times {\bf T}^{N-1})/ {\cal S}_{N}$ -
just the moduli space of flat connections on
the torus. Here ${\bf T}^{N-1}$ is a Cartan torus for
$SU(N)$ (center of mass is fixed) and ${\cal S}_{N}$
is a symmetric group.
We call them fermions,
because  their wavefunction vanishes on the diagonals $z_{i} = z_{j}$
in the $N-1$-torus (it contains a factor like
$\Delta ( e^{{\sf i}\h_{i}} )$ - a group-like Vandermonde).
It is obvious,
that
the corresponding Schr\"odinger equation will be a difference
equation, because
$cos( \frac{2\pi{\sf i}}{{\kappa} + N} {\partial}_{q})$
is a finite difference operator.
Its eigenfunctions on the circle are
exponents $e^{2 {\pi}{\sf i}n q}$ ($n$ is necessarily integral) with
eigenvalue $$ E_{n} = cos (\frac{2\pi n}{{\kappa} +N}) $$ and the total
spectrum will be given by the sum $$ \sum_{i} E_{n_{i}} $$ where we have to
impose the following conditions, coming from the invariance $E_{n} = E_{n +
{\kappa} + N}$ and the symmetric group action:  \beq ({\kappa} + N ) >
n_{N}> \dots > n_{i} > \dots > n_{1} \geq 0 \label{restr1} \eeq \beq
\sum_{i} n_{i} < ({\kappa} + N)
\label{restr2}
\eeq

In our situation fermions on the circle are not
completely "free" and the spectrum
is restricted. For our purposes will be more
convinient to form a vector
$\hh = (n_{i} - i), i=1,2,\dots,N$. It will correspond to the
dominant weights of the $SU(N)$ irreducible representations.
Now we can write an answer for the  spectrum:
$$
E_{\hh} = \sum_{i} cos(\frac{2\pi \hh_{i}}{\kappa + N})
$$

and $\hh$ satisfy the selection rule: there exists non-trivial
intertwiner

$$
\Phi_{\hh} : \alpha_{\hh} \to \alpha_{\hh} \otimes R_{\nu}
$$

(essentually it implies, that $\hh = \lambda + \nu \rho$, where
$\lambda_{i} \geq 0, i=1, \dots, N$.

An explanation of
this spectrum can be
derived from the properties of gauged $G/G$ WZW model \cite{bt + ger}.
Namely,
to get a
partition
function of
our interest
we have to compute a path integral on the two-torus:
\beq
\int
{\ddd}A{\ddd}g
\exp(
S_{GWZW}(g, A) +
\int_{T^{2}} {\sf Tr} ( g + g^{-1}) )\chi_{R_{\nu}} ( P \exp \int_{C} A )
\label{masterpath}
\eeq
where
$C$ is a
non-contractible
contour on the torus,
$\chi_{R_{\nu}}$
is a character in
the representation $R_{\nu}$.
By the standard arguments we can calculate
it by
cutting the torus along
$C$ and computing first
the path integral in the theory
$S_{GWZW}(g, A) + \int {\sf Tr} ( g + g^{-1}) $
on the annulus with coinciding monodromies $h$ of
the restrictions of
the
connection
$A$ on
the boundaries
of the annulus and
then integrate the answer over $H$ with the weight $\chi_{R_{\nu}}(h)$.
To make a
further
progress one needs some information about the path integral on the annulus
(or cylinder, what is the same). In principle, this can be deduced from
\cite{bt + ger}.
We will not present here the whole derivation, just note that
the path integral on
the cylinder can be localized
(either by some
equivariant cohomology-like
technique for $SU(N)$ gauge group itself or by using a gauge, where
$g$ is diagonal and then use the
usual localization for abelian gauge theory
\cite{bt + ger}) on the $g$ with special values of
eigenvalues,
namely those which correspond to
the integrable representations of
$SU(N)_{\kappa}$,
i.e. with the signature,
given by numbers $n_{i}$, satisfying the same restrictions
(\ref{restr1}),(\ref{restr2}).
  In arbitrary gauge theory in two dimensions the
path integral on the disk and
on the cylinder can be written in terms of sum over all
irreducible representations of the gauge group via the characters
in these representations. We conjecture that taking into
account Wilson line in the representation $R_{\nu}$ will lead
us to Macdonald polynomials \cite{ek}. We hope to return to
more detailed investigation in the future.

\section{Conclusions and discussion}

In this paper  we have
considered some particular integrable deformations of Calogero-Moser systems,
which preserve their trigonometric form,
but make the quantum Schr\"odinger operator a finite difference operator.
We have observed an equivalence of this theory with a gauged
$SU(N)/SU(N)$ WZW model with
Hamiltonian on the cylinder with inserted Wilson line.
The level of WZW model
and the representation in which Wilson line is taken are encoded in
the coupling constant of the model and in the mass scale.

There are some important and intriguing questions,
which remained beyond the scope of this paper.
The first one is the elliptic analogue of the topological theories
like gauged WZW or its
degeneration - Yang-Mills theory.
We expect this elliptic generalization to exist,
 because in the corresponding
quantum integrable systems of particles there is such an
analogue\footnote{See \cite{gorell} for the progress in this direction}.
When elliptic curve degenerates
one gets trigonometric version of integrable model.
One of the possible ways is to consider double-loop algebras
\cite{morozov},\cite{ef} and/or
Sklyanin algebra and its multivariable generalizations
as the algebras of symmetries of corresponding
theories.  Some indications of the appearence of Sklyanin algebra are in the
Askey-Wilson polynomials, which are particular examples of the
wave-functions of systems under consideration \cite{ruu},\cite{zabrodin}.

Another interesting question concerns investigation of different limits
of our theories when coupling constants go to some extremal values.
In fact, we have two important parameters in hands:
$$
q = e^{\frac{2\pi i}{\kappa + N}}
$$
and
$$
t = q^{\nu + 1}
$$

We have conjectured that in the gauged $G/G$ WZW theory with the
Wilson line in the representation $R_{\nu}$ the answers
for path integrals on the surfaces with boundaries can be expressed
in terms of Macdonald polynomials $P_{\hh} ( x ; q,t)$. These polynomials
have very interesting limits (for example,
they are responsible for the zonal spherical
functions on the p -adic groups,namely Mautner-Cartier polynomials when
$q=0,t=1/p$) if $q,t$ tend to some exceptional values.  Unfortunately, in
our approach $q$ should be a root of unity and $t$ should be an integral
power of $q$, so there is no obvious way how to recognize all this beauty of
Macdonald polynomials in gauged $G/G$ WZW theory for $SU(N)$
group. Presumably
$SL(N,C)$ can overcome this problem.

It is known that Ruijsenaars system is closely related with the solitonic
sector of Toda like theories,in particular asymptotics of the wave funstions
provides S-matrix for scattering in the solitonic sector \cite{solitons}.
The qualitative explanation of this relation looks as follows. Lagrangian
approach for Toda theories is based on GWZW actions with some additional
dependence on the spectral parameter. In the consideration above we added
Hamiltonians and choosed some peculiar  level of the momentum map. Thus
the final action is nothing but the generating functional for GWZW while
the level of the momentum map fixes the particular solitonic solutions.
Hence we considered in essence the generating functional and solitonic
S-matrix naturally appeares in this context. Certainly this relation needs
for further clarification, especially in context of (q)KZ equations
describing the soliton scattering amplitudes. We plan to consider this
problem elsewhere.

Recently, in the works of Dunkl, Opdam and Heckmann \cite{dunkl} a powerful
method of
solving Sutherland and Calogero - type models was found. It is based on the
introducing of operators acting on the group algebra of the symmetric group
${\bf C}[{\cal S}_{N}]$. This method was applied also in a seria of papers to
a more general systems, including spin long-range interactions (see, e.g.
\cite{bghp+fv+bhv}).
To obtain a meaning of these operators
in a gauge theory remains to be quite important question. They
could be related
somehow to the problem of Gribov copies in a gauge fixing in a gauge theory
and, therefore can be of some improtance in a four-dimensional gauge theory.

Another important direction is
the finite-dimensional verification of equivalences, obtained between
finite-dimensional system and (at first sight) infinite-dimensional.
In the case of Sutherland model one can check, that there is an intermediate
step in the procedure of Hamiltonian reduction, leaving us with the cotangent
bundle to the group $SU(N)$. In the case of affine group instead of affine Lie
algebra the situation seems to be slightly complicated and
one hopes to get as a finite-dimensional analogue of
$T^{*}SU(N)$ one of the $SU(N)$ doubles, most probably the Heisenberg double,
which has an open dense  symplectic leaf.
But, instead of Hamiltonian action of
$SU(N)$ on its cotangent bundle in this situation we have a
Poissonian action of Poisson-Lie group,
which is slightly more involved situation \cite{fro}, \cite{almal}.
Still it seems to be interesting question to  answer. We think
that in this way one can understand the relation
of the quantum group representation theory to our problems \cite{ek}.

Let us finally note that finite-dimensional Ruijsenaars system can be
consistently quantized.To this end one should find a proper normal ordered
form od the Hamiltonian.For the $N=2$ case it has the following form

$$
 H=\Phi_{+} T_{-} \Phi_{-} +\Phi_{-}T_{+} \Phi_{+}
$$
$$
T_{\pm}\Psi(x) =\Psi (x{ \pm}\frac{1}{k+N})
$$

where $\Phi_{+}= \Phi_{1} ,\Phi_{-}=\Phi_{2}$.In our approuch we have
the very normal ordering which preserves the integrability.

\section{ Acknowledgements}
We are grateful to A.Alekseev, V.Fock, A.Malkin, A.Marshakov,\\
    M.A.Olshanetsky,  A.Rosly,  S.Shatashvili  for valuable discussions.
N.N. expresses his grattitude to  Uppsala University and especially
to Prof. A.Niemi for kind hospitality
while a part of this work was done. We are also grateful to Aspen Center
for Physics where the paper was partially prepared.
Research of N.N. was supported in part by Soros Foundation Grant
awarded by American Physical Society, and by Grant of
French Academy of Sciences.
We also acknowledge a support of RFFI, grant \# 93-02-14365

\bigskip

\appendix{\Large Appendix A}

\setcounter{equation}{0}

\bigskip

In this section we prove the relation (\ref{Lax}).

The necessary condition for the equation (\ref{commutant})
to hold is that the eigenvalues
 of matrices $Z$ and
$$
\exp(J) Z = \lll (Id + (\lll^{-N} -1)f \otimes f^{+}) Z
$$
coincide.

It leads to the system of
equations

\beq
1+(\lll^{-N}-1) \sum_{i=1}^{N}
\frac{\Phi_{i}}{1-\frac{z_{i}\lll}{z_{k}}} = 0,
\label{cpneq}
\eeq

$$
k = 1, \dots, N ; \; \Phi_{i} = | f_{i} |^{2}
$$

Also we have to satisfy the normalization condition $\sum_{i} \Phi_{i} = 1$.
Let us note that our equations (\ref{cpneq})
resemble those arising
in the definition of the Gelfand-Tzetlin coordinates \cite{samitz}.
The reason is obvious:
Gelfand-Tzetlin coordinates
are related to the imbeddings of
$S(U(N-1) \times U(1))$ into $SU(N)$,
the quotient $SU(N)/ S (U(N-1) \times U(1))$ is precisely
${\bf CP}^{N-1}$ - our coadjoint orbit.
Also let us note that the solution of these equation
(written on a next line) looks like expression in
the right hand side of the Bethe ansatz equations -
the reason for this is to be understood.
Our claim is: $\Phi_{i} = Q^{+}(z_{i})$.
 We prove this formula
by twofold evaluation of
the following contour integrals in the complex plane around
infinity:

$$
I_{0} = \frac{1}{2\pi i} \oint_{C_{\infty}} dz \frac{P({\lll}z)}{z P(z)}
$$

and

$$
I_{k} = \frac{1}{2\pi i}
\oint_{C_{\infty}} dz \frac{P({\lll}z)}{(z - {\lll}^{-1} z_{k}) P(z)}
$$

Contour $C_{\infty}$ is oriented counterclockwise and goes around infinity.
Both of them are obviously equal to $\lll^{N}$ (residue at infinity).
On the other hand, $I_{0}, I_{k}$
are given by the sum of residues at poles of the integrands inside of
$C_{\infty}$. It gives

$$
I_{0} = 1 + \sum_{i} ({\lll}^{N}-1) Q^{+}(z_{i})
$$

$$
I_{k} = \sum_{i} Q^{+}(z_{i})
\frac{{\lll}^{N}-1}{1-\frac{z_{k}}{z_{i}} \lll^{-1}}
$$

and we obtain j
This is equivalent to
$\lll (Z v_{i} + (\lll^{-N} -1) <f, Z v_{i}> f) = z_{i} v_{i}$, which leads
to

$$
v_{i} = g_{i} (\lll^{-N} -1) (z_{i} \lll^{-1} - Z)^{-1} f
$$

where $g_{i}$ are some constants to be found from the normalization
$$
<v_{i}, v_{j}> = \delta_{ij}
$$
To this end we have to calculate an
expression

\beq
{\Psi_{i}}^{-1} = |g_{i}|^{-2}= (2-\lll^{N}-\lll^{-N}) \sum_{j}
\frac{\Phi_{j}}{(z_{i}\lll^{-1}-z_{j})(z_{i}^{-1}\lll - z_{j}^{-1})}
\label{cpneq2}
\eeq

This expression can be evaluated with help of residues of the following
contour integral

$$
J_{i} = \frac{1}{2\pi {\sf i}} \oint_{C_{\infty}} \frac{P(\lll z)}
{(z-z_{i}\lll^{-1})^{2}P(z)}
$$

It is obvious that $J_{k} = 0$, because at infinity the integrand behaves  as
$ z^{- 2}$ . On the other  hand, it implies:

\beq
\frac{\lll P^{\prime}(z_{k})}{P(\lll^{-1} z_{k})} = -
\sum_{i} Res_{z=z_{i}}\frac{P(\lll z)}{(z-z_{i}\lll^{-1})^{2}P(z)},
\label{res}
\eeq
thus,
$$
\Psi_{i} = Q^{-}(z_{i})
$$

Now we are in a position to write down an expression  for matrix
$\tilde G$ :
$$
{\tilde G}_{ij} = e^{i\theta_{i}}{\lll}^{\frac{N-1}{2}}
\frac{{\lll}^{-N}-1}{z_{i}- \lll^{-1}z_{j}} (\Psi_{i} \Phi_{j})^{1/2}
$$

Finally, after substituting here (\ref{cpneq2}),(\ref{res}) we
just get (\ref{Lax}). The phases $\theta_{i}$
are restricted according to the condition $det({\tilde G}) = 1$,
so this is a condition, imposed on the $\sum_{i} \theta_{i}$,
which doesn't take us here. Still, for completeness,
let us calculate this determinant:

$$
det ({\tilde G}) = \sum_{\sigma \in S_{N}} (-)^{\sigma}
\prod_{i} {\tilde G}_{i, \sigma(i)} =
$$

$$
 = e^{ i\sum_{j}{\h_{j}}}{\lll}^{\frac{N-1}{2}}
\prod_{j} (Q^{+}(z_{j})Q^{-}(z_{j}))^{1/2}
\sum_{\sigma \in S_{N}} (-)^{\sigma}
\prod_{i} \frac{{\lll}^{-N}-1}{{\lll}^{-1}z_{i} - z_{\sigma (i)}} =
$$

$$
 = e^{ i\sum_{j}{\h_{j}}}{\lll}^{\frac{N-1}{2}}
({\lll}^{-N}-1)^{N}
{\left[ \frac{({\lll}-1)
({\lll}^{-1}-1)}{({\lll}^{N}-1)({\lll}^{-N}-1)} \right]}^{1/2}\times
$$

$$
\times\prod_{j} (Q^{+}(z_{j})Q^{-}(z_{j}))^{1/2}
det || \frac{1}{z_{i} - {\lll}^{-1}z_{j}} || =
$$
\centerline{\em (Cauchy identity)}
$$
= e^{ i\sum_{j}{\h_{j}}} {\lll}^{\frac{N-1}{2}}
\prod_{j} \frac{{\lll}^{-N}-1}{z_{j} (1-{\lll}^{-1})}
\prod_{j<k} \frac{(z_{j}-z_{k})^{2}}{(z_{j}-z_{k}{\lll}^{-1})
(z_{j}{\lll}^{-1}-z_{k})}\times
$$

$$
\times \frac{({\lll}z_{j}-z_{k})({\lll}^{-1}z_{j}-z_{k})}
{(z_{j}-z_{k})^{2}} =
$$

$$
= e^{i\sum_{j} \h_{j}} = 1
$$
The determinant $det || \frac{1}{z_{i} - {\lll}^{-1}z_{j}} ||$
can be evaluated by representing
it as a correlator of free fermions on a complex
plane inserted into points
$z_{i}$ and ${\lll}z_{i}$ -
it is a "physical"
proof of Cauchy identities used also in
\cite{ruu}\footnote{We thank
A.Marshakov for pointing out
this useful application of Wick theorem}.
So, the answer
for this
determinant cancels precisely all
$z_{i}$ dependent factors in the product (this
could be also
deduced from the absence of poles of this meromorphic
function, therefore, it should be constant).
\bigskip

\appendix{\Large Appendix B}

\setcounter{equation}{0}

\bigskip

Here we derive an expression for the wavefunctions of Calogero model via the
large ${\kappa}$ limit in those for Sutherland model. We start from the
expression
for the transition kernel

$$
{\cal K} (g_{2}, v_{2} | g_{1}, v_{1} ) = \sum_{\alpha} \exp( -
\frac{T}{{\kappa}^{2}} Q (\alpha )) {\cal K}_{\alpha} (g_{2}, v_{2} | g_{1},
v_{1} )
$$

\beq
{\cal K}_{\alpha} (g_{2}, v_{2} | g_{1}, v_{1} ) =
 \int_{SU(N)} dh \chi_{\alpha} ( h g_{1} h^{-1} g_{2}^{-1}) <v_{2}|
T_{R_{\nu}}(h) |v_{1}>
\label{amplit}
\eeq

Let us write $g_{1,2} = \exp ( \frac{i}{\kappa} Q_{1,2})$, then
$ h g_{1} h^{-1} g_{2}^{-1} = \exp ( \frac{i}{\kappa} \Phi )$ with

$$
\Phi = h Q_{1} h^{-1} - Q_{2} + {\cal O}(\frac{1}{\kappa})
$$

Now we can write down an expression for the eigenvalues of the matrix $h g_{1}
h^{-1} g_{2}^{-1}$ as $e^{\frac{i\Phi_{j}}{\kappa}}$ where $\Phi_{j}$ are
eigenvalues of $\Phi$. By the Weyl character formula we have:

\begin{eqnarray}
&& dim ( \alpha) = \frac {\Delta (\ppb)}{\Delta (i)}
\nonumber\\
&& \chi_{\alpha}(U) = \frac{det||{\beta}_{j}^{\ppb}||}{{\Delta} (\beta_{j})}
\label{irreps}
\end{eqnarray}
where $U $ is conjugate to $diag(\beta_{1},\ldots,\beta_{N})$.
Here  collection of $\ppb$'s is the highest weight $\hh$ of the
irreps $\alpha_{\hh}$.

The formula we
will derive uses the representation of the large $\kappa$ limit
of
the normalized character through
the Harish-Chandra(-Itsykson-Zuber) integral,
namely, if  $\frac{\hh}{\kappa} = {\bar P}$   is fixed, then

$$
\lim_{{\kappa} \rw { \infty}}\frac{\chi_{\alpha}(U)}{dim(\alpha)} =
\int_{SU(N)} dh \exp ( i Tr ({\bar P} h{\bar Q}h^{-1})
$$
where $\exp (i\frac{\bar Q}{\kappa})  = diag ( U )$, and $diag (U) $ denotes
any diagonalized form of the matrix $U$ . Substituting this expression into
the integral in (\ref{amplit}), we get an integral of the type:

\begin{eqnarray}
&& {\cal K}_{\bar P} (g_{2}, v_{2} | g_{1}, v_{1} ) =
\nonumber\\
&& \Delta ({\bar P}) {\kappa}^{N(N-1)/2} \times
\nonumber\\
&& \int\int_{SU(N)}dh dh^{\prime} \exp ( i Tr (h^{\prime} (h Q_{1} h^{-1} -
Q_{2}) {h^{\prime}}^{-1} {\bar P})
\nonumber\\
&& <v_{2}| T_{R_{\nu}}(h) |v_{1}>
\label{limamplit}
\end{eqnarray}
Finally, we can extract a wavefunction from the condition:

$$
{\Psi}_{\bar P}^{\dag} ({\bar Q}_{2}) {\Psi}_{\bar P} ({\bar Q}_{1}) =
\int\int_{SU(N)} dh dh^{\prime} {\cal K}_{\bar P} (g_{2}^{h}, v_{2}^{h} |
g_{1}^{h^{\prime}}, v_{1}^{h^{\prime}} )
$$
where $g^{h} = h g h^{-1}$, $v^{h} = T_{R_{\nu}}(h)v$.

It yields essentually :

$$
{\Delta}({\bar P}) {\Delta}({\bar Q}) \int_{SU(N)} dh \exp ( i tr
({\bar P} h {\bar Q} h^{-1}) <0| T_{R_{\nu}} (h) |0>
$$

   In the case of $N=2$
this integral reduces to the integral representation for the Bessel
function.  It is interesting to note, that the initial expression of the
wavefunction for the Sutherland model could be regarded as the same formula
applied to the central extended loop group -- in that case we would get some
path integral, which we can evaluate using, for example, localization
technique, or, by choosing appropriate coordinates on the co-adjoint orbit
of affine algebra (\cite{samson}).

\end{document}